\documentclass[aps,prd,twocolumn,superscriptaddress,preprintnumbers,floatfix,nofootinbib,notitlepage,showkeys,showpacs]{revtex4-2}
\usepackage[utf8]{inputenc}
\usepackage{epsfig,amsfonts,amsthm}
\usepackage{epstopdf}
\usepackage{amsmath,amssymb}
\usepackage{latexsym}
\usepackage{bbm}
\usepackage{array}
\usepackage{comment}
\usepackage{hyperref}

\usepackage{color}
\usepackage{pb-diagram}
\usepackage{slashed}
\usepackage{cancel}
\newcommand{\be}{\begin{equation}}
\newcommand{\ee}{\end{equation}}
\newcommand{\bea}{\begin{eqnarray}}
\newcommand{\eea}{\end{eqnarray}}

\newcommand{\bvec}[1]{{\bf{#1}}}

\usepackage{color}
\definecolor{myorange}{rgb}{1,0.5,0}

\begin{document}
\bibliographystyle{OurBibTeX}


\title{Solar neutrinos and dark matter detection with diurnal modulation}

\author{Sebastian Sassi}
\email{sebastian.k.sassi@helsinki.fi}
\affiliation{Department of Physics, University of Helsinki, 
	P.O.Box 64, FI-00014 University of Helsinki, Finland}
\affiliation{Helsinki Institute of Physics, 
	P.O.Box 64, FI-00014 University of Helsinki, Finland}

\author{Abolfazl Dinmohammadi}
\email{dinmohammadi@znu.ac.ir}
\affiliation{Department of Physics, Faculty of Science, University of Zanjan,  P.O. Box 45195-313, Zanjan, Iran}

\author{Matti Heikinheimo}
\email{matti.heikinheimo@helsinki.fi}
\affiliation{Department of Physics, University of Helsinki, 
                      P.O.Box 64, FI-00014 University of Helsinki, Finland}
\affiliation{Helsinki Institute of Physics, 
                      P.O.Box 64, FI-00014 University of Helsinki, Finland}

\author{Nader Mirabolfathi}
\email{mirabolfathi@physics.tamu.edu}
\affiliation{Department of Physics and Astronomy, Texas A\& M University}

\author{Kai Nordlund}
\email{kai.nordlund@helsinki.fi}
\affiliation{Department of Physics, University of Helsinki, 
                      P.O.Box 64, FI-00014 University of Helsinki, Finland}
\affiliation{Helsinki Institute of Physics, 
                      P.O.Box 64, FI-00014 University of Helsinki, Finland}
\author{Hossein Safari}
\email{safari@znu.ac.ir}
\affiliation{Department of Physics, Faculty of Science, University of Zanjan, P.O. Box 45195-313, Zanjan, Iran}
                  
\author{Kimmo Tuominen}
\email{kimmo.i.tuominen@helsinki.fi}
\affiliation{Department of Physics, University of Helsinki, 
                      P.O.Box 64, FI-00014 University of Helsinki, Finland}
\affiliation{Helsinki Institute of Physics, 
                      P.O.Box 64, FI-00014 University of Helsinki, Finland}

\begin{abstract}
\noindent
{We investigate the diurnal modulation of the event rate for dark matter
scattering on solid targets arising from the directionally dependent
defect creation threshold energy. In particular, we quantify how this
effect would help in separating dark matter signal from the neutrino background.
We perform a benchmark analysis for a germanium detector and compute how the
reach of the experiment is affected by including the timing information of the scattering events.
We observe that for light dark matter just above the detection threshold the magnitude of the annual modulation is enhanced. In this mass range using either the annual or diurnal modulation information provides a similar gain in the reach of the experiment, while the additional reach from using both effects remains modest.
Furthermore, we demonstrate that if the background contains a feature exhibiting an annual
modulation similar to the one observed by DAMA experiment, the diurnal modulation provides for an additional handle to separate dark matter
signal from the background.
}
 \end{abstract}
\preprint{HIP-2021-3/TH}

\maketitle

\section{Introduction}

As the exclusion limits for direct detection of dark matter (DM) particles in the traditional
WIMP mass range are approaching the neutrino floor, more focus is being shifted
towards low-mass dark matter with $m_{\rm DM}\ll 10$ GeV, where the recoil
kinematics forbid efficient detection in, e.g., a liquid xenon target.
Ionization or phonon-mediated solid state detectors are currently the technologies of choice for very low mass DM searches. Many technologies are  currently offering detection thresholds that are sensitive to DM--nucleus elastic scattering down to single-electronic excitation ($\sim 10$ eV) thresholds ~\cite{Single, sensei, edelweiss}. Although those technologies are currently demonstrated with low-mass modules $\sim$ g, significant strides have been made to scale up the detector masses to $\sim$ kg~\cite{CFGe, CFSi}.   

Compared to the isotropic liquid target, a solid material exhibits an additional
feature due to the anisotropy of the crystal lattice. Therefore the nuclear scattering
process shows a directional dependence at low energies. Specifically, the threshold
energy for defect creation varies significantly as a function of the recoil direction.
This leads to a diurnal modulation in the ionization signal that depends on the direction of the flux of
the scattering particles, as demonstrated in~\cite{Kadribasic:2017obi}. As a threshold
effect, the diurnal modulation signal is expected to be observable only within a
relatively narrow interval of DM masses. For example, for germanium
this interval extends over a few hundred MeV above $m_{\rm{ DM}}\simeq 300$ MeV.
However, using different detector materials allows variation of the threshold
energy and consequently will allow probing of a wider range of DM masses~\cite{Heikinheimo:2019lwg}. Still at lower DM masses, corresponding to recoil energies well below the defect creation threshold, the anisotropy of the crystal lattice gives rise to diurnal modulation in phonon and electron excitations, as discussed in~\cite{Coskuner:2019odd,Trickle:2019nya,Coskuner:2021qxo}.

In this paper we provide a more elaborate analysis of
the potential gain due to this anisotropic threshold effect in the sensitivity reach beyond the solar neutrino coherent scattering backgrounds.
We will show that for a generic parametric model
containing both annual and diurnal modulation features, the gain in the sensitivity reach of the experiment strongly depends on the relative amplitudes of the former versus the latter.
For the particular case study of a germanium target, since both modulation amplitudes are of the same order,  we show that the improvement of the reach of the experiment below the neutrino floor is
saturated by using either of the modulation signals, and additional gain from using both is only obtained for a prohibitively large exposure.
This is because the energy scale for the threshold anisotropy (diurnal modulation) is of the same order as the DM energy modulation due to the seasonal variation of the velocity of Earth with respect to the DM halo (annual modulation).
However, as the systematics of the experiment might be difficult to control over long periods of data taking, and could even induce a spurious modulation signal~\cite{Buttazzo:2020bto, Messina:2020pnt}, the redundancy provided by the diurnal modulation feature demonstrated here should allow for improved confidence in the DM origin of the signal.

As a further case study we consider the case of a hypothetical annually modulating
background. The DAMA/LIBRA experiment~\cite{Bernabei:2018jrt} has observed a clear signature of an
annual modulation compatible with the expectation from the motion of Earth with
respect to the DM background. However, the DM interpretation of this observation
seems to be ruled out by the incompatibility of the DAMA result with other
DM direct detection experiments (e.g., \cite{Aprile:2018dbl, Akerib:2016vxi, Wang:2020coa}) which have put firm exclusion constraints over
the mass range favored by DAMA. Therefore, it is plausible to take the DAMA
result as a signal of an unknown background feature whose modulation coincides
with that of the DM background. In such case we demonstrate that the diurnal
modulation offers an additional quantitatively important criterion to separate
the DM signal from the background.

The paper is organized as follows: In section~\ref{sec:DMeventrate} we review the
computation of the DM event rate in the presence of directionally dependent
energy threshold and in section~\ref{sec:nueventrate} we extend these results
to the rate of solar neutrinos. In section~\ref{sec:likelihood} we formulate
the likelihood analysis which we will then apply to obtain results which we
present and discuss in detail in section~\ref{sec:results}. In section~\ref{sec:checkout}
we present our conclusions and outlook.

\section{DM event rate}
\label{sec:DMeventrate}

The computation of the event rate, taking into account the directional dependence of the threshold energy for defect creation in germanium, has been discussed in \cite{Kadribasic:2017obi, Heikinheimo:2019lwg}, and here we only outline the key steps of the computation. The DM--nucleus scattering event rate (in case of no velocity dependence in the DM--nucleus coupling) is given by
\be
\frac{dR}{dEd\Omega_q}= \frac{\rho_0}{4\pi m_\chi}\frac{|\mathcal{M}_N|^2}{16\pi m_N^2 m_\chi^2}\hat{f}(v_{\rm min},\hat{\bvec{q}}),
\label{DM-nucleus rate}
\ee
where $m_\chi$ is the DM mass, $m_N$ is the mass of the target nucleus, $\mathcal{M}_N$ is the DM--nucleus scattering amplitude, $\rho_0$ is the local DM energy density, and $\hat{f}$ is the Radon transform of the DM velocity distribution $f(v)$, defined as
\be
\hat{f}(v_{\rm min},\hat{\bvec{q}}) = \int f(v)\delta(\bvec{v}\cdot\hat{\bvec{q}}-v_{\rm min})\,d^3v.
\ee
Here $v_{\rm min} = \sqrt{m_N E/(2\mu_{\chi N}^2)}$ is the minimum velocity of the DM particle in the lab frame required to exite a nuclear recoil with energy $E$ and $\mu_{\chi N} = m_\chi m_N/(m_\chi+m_N)$ is the reduced mass of the DM--nucleus system.

The DM--nucleus scattering amplitude is related to the DM--nucleon amplitude $\mathcal{M}_n$ via
\be
|\mathcal{M}_N|^2 = A^2F(E)\frac{m_N^2}{m_n^2}|\mathcal{M}_n|^2,
\ee
where $F(E)$ is the nuclear form factor, $A$ is the mass-number of the nucleus and $m_n$ is the mass of the nucleon. The spin-independent DM--nucleon scattering cross section is defined as
\be
\sigma_{\chi n} = \frac{|\mathcal{M}_n|^2}{16\pi(m_\chi+m_n)^2}.
\ee
Therefore we can express the DM-nucleus scattering rate (\ref{DM-nucleus rate}) in terms of the spin-independent DM--nucleon cross section as
\be
\frac{dR}{dEd\Omega_q}=\frac{\rho_0 A^2 F(E)\sigma_{\chi n}}{4\pi m_\chi \mu_{\chi n}^2}\hat{f}(v_{\rm min},\hat{\bvec{q}}),
\ee
where $\mu_{\chi n} = m_\chi m_n/(m_\chi+m_n)$ is the reduced mass of the DM--nucleon system and $\hat f$ is the Radon transform of the DM velocity distribution.

\begin{figure*}
	\begin{center}
		\includegraphics[width=0.95\linewidth]{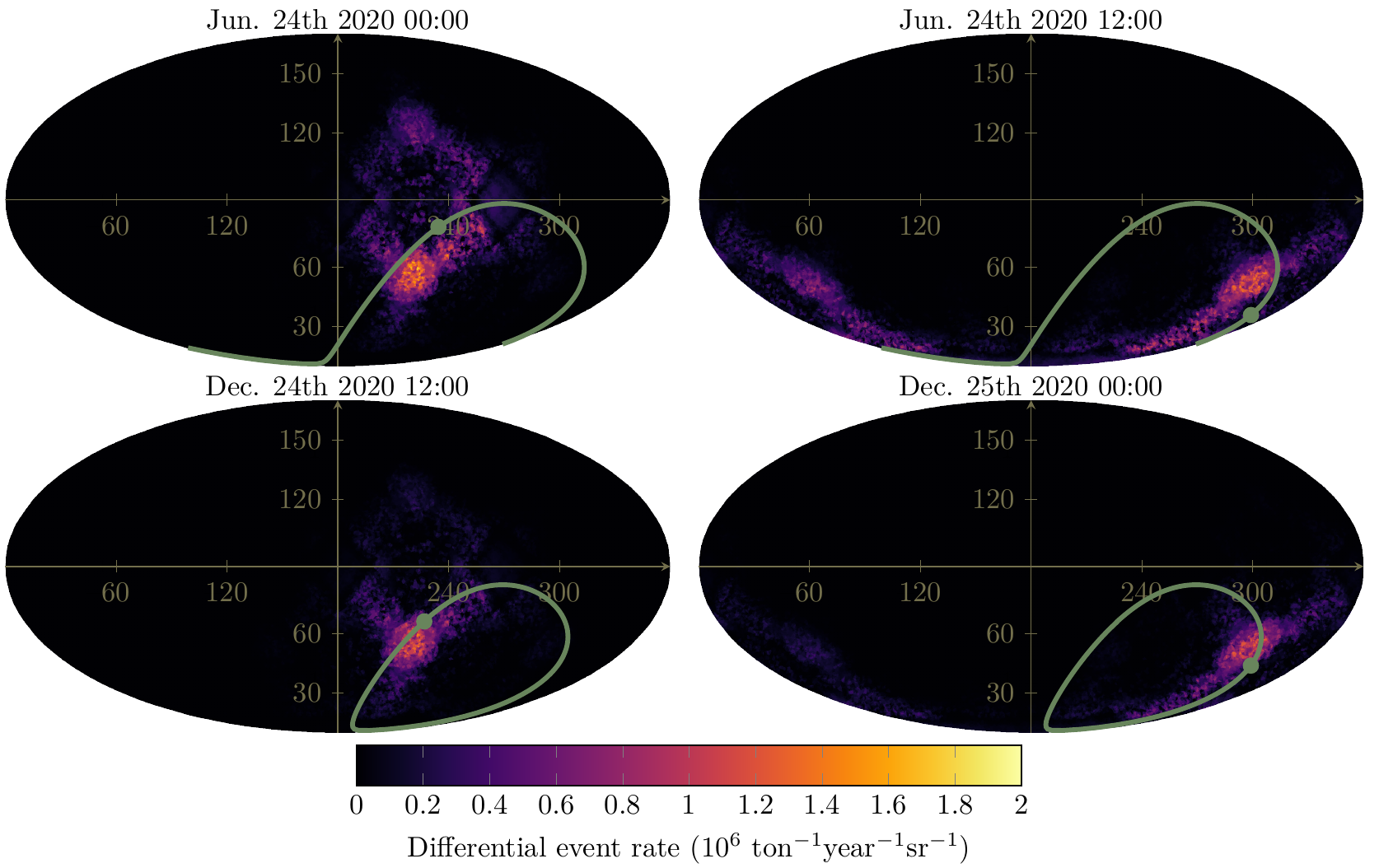}
		\caption{The recoil event rate for 400 MeV DM particle with a $10^{-39}\ \text{cm}^2$ cross section as a function of recoil direction for germanium in June and December, during the moments of daily maxima. The dot in each figure shows the direction of the DM wind in the lab frame at the moment, and the loop shows how the direction varies during one day.}
		\label{directional_rates}
	\end{center}
\end{figure*}

In this work we use the standard halo model (SHM) for the DM velocity distribution, for which the analytical formulas of the Radon transform, integrated over recoil energy, can be found in~\cite{Heikinheimo:2019lwg}. The integral over the solid angle is computed as a Monte Carlo sum over the recoil directions, with the threshold energy for each sampled direction substituted as the lower limit for the energy integral, as discussed in more detail in \cite{Heikinheimo:2019lwg}. The resulting event rate is a function of time, due to the changing direction and amplitude of the dark matter velocity vector in the lab frame. This effect is demonstrated in figure \ref{directional_rates}, where the instantaneous event rate for a 400 MeV DM particle is shown during the daily maxima, for two dates, corresponding to the minimum (December) and maximum (June) amplitudes of the DM wind speed in the lab-frame. Notice that the times of day for the phases of the velocity vector are shifted by approximately twelve hours between the two dates because its period is one sidereal day and not a solar day.

\section{Neutrino event rate}
\label{sec:nueventrate}

\begin{table}
\center
\begin{tabular}{lll}
\\\hline
                        & $E_\text{max}$ (MeV)  & $\Phi_0$ ($\text{cm}^{-2}\text{s}^{-1}$)  \\\hline
$\Phi(\mathrm{pp})$     & 0.423                 & $5.98(1\pm 0.006)\cdot 10^{10}$           \\
$\Phi(\mathrm{pep})$    & 1.445                 & $1.44(1\pm 0.01)\cdot 10^8$               \\
$\Phi(^7\mathrm{Be})$   & 0.386, 0.863          & $4.93(1\pm 0.06)\cdot 10^9$               \\
$\Phi(^8\mathrm{B})$    & 15.1                  & $5.46(1\pm 0.12)\cdot 10^6$               \\
$\Phi(^{13}\mathrm{N})$ & 1.198                 & $2.78(1\pm 0.15)\cdot 10^8$               \\
$\Phi(^{15}\mathrm{O})$ & 1.732                 & $2.05(1\pm 0.17)\cdot 10^8$               \\
$\Phi(^{17}\mathrm{F})$ & 1.736                 & $5.29(1\pm 0.20)\cdot 10^6$               \\\hline
\end{tabular}
\caption{Maximum neutrino energies and fluxes of the solar neutrinos used from the B16-GS98 solar model \cite{Vinyoles:2016djt}.}
\label{tab:fluxes}
\end{table}

\begin{figure*}
\begin{center}
\includegraphics[width = 0.45\textwidth]{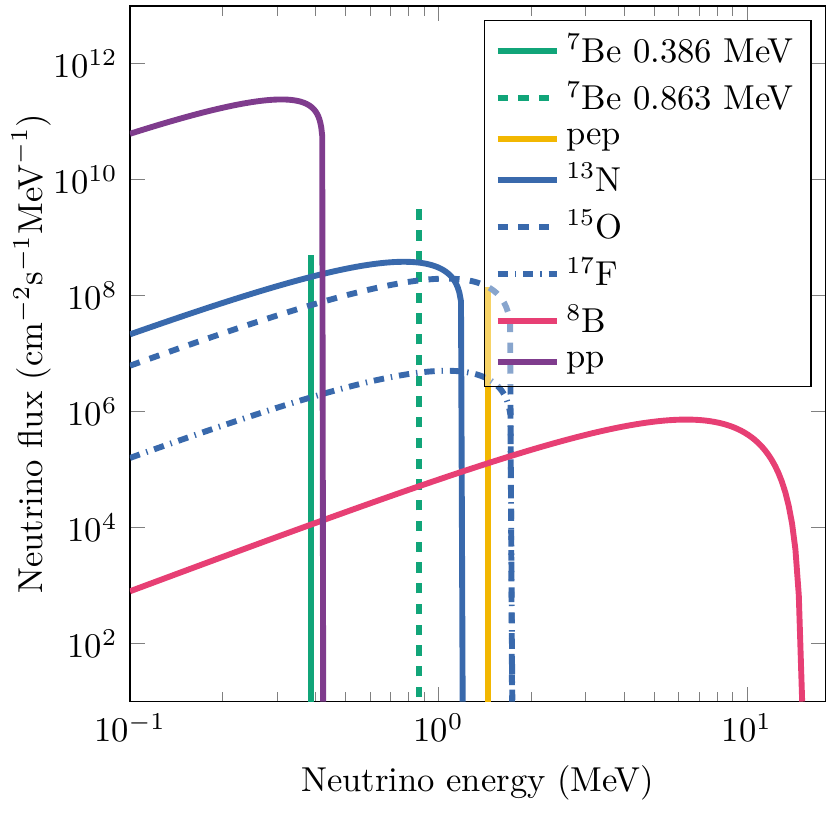}
\includegraphics[width = 0.45\textwidth]{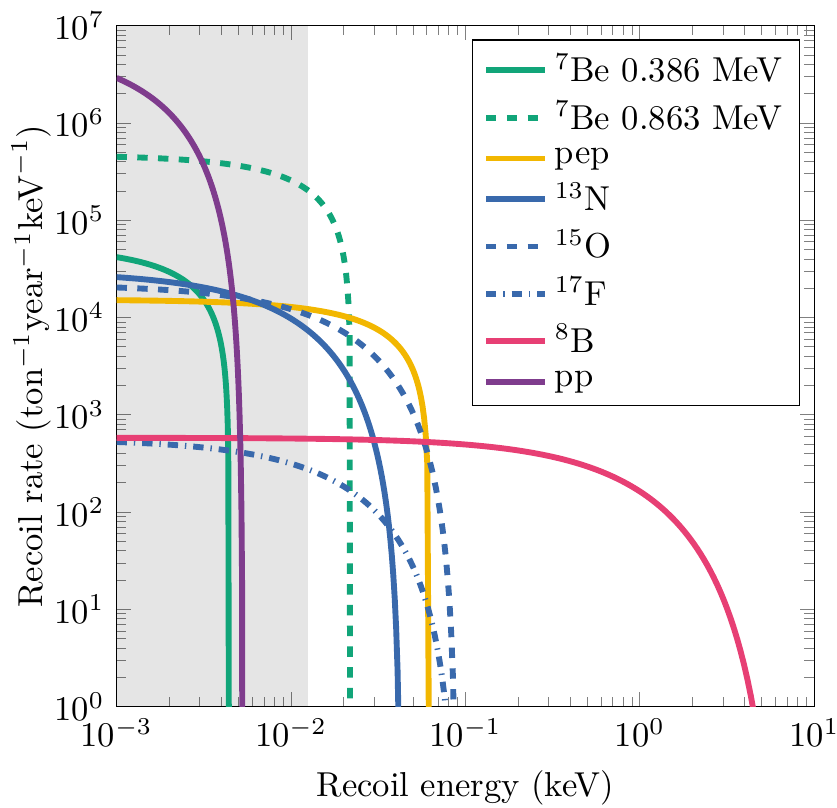}
\caption{Left: Neutrino energy spectrum $d\Phi/dE_\nu$. Right: Recoil spectrum in germanium for the relevant components of the solar neutrino flux. The shaded region corresponds to $E_r<12.5$ eV which is the minimum defect creation threshold energy in germanium.}
\label{nuflux}
\end{center}
\end{figure*}

To determine the neutrino event rate, we follow the analysis presented in \cite{OHare:2015utx}.
The left panel of figure \ref{nuflux} shows the solar neutrino flux $d\Phi/dE_\nu$ for the relevant components.
The differential neutrino-nucleus scattering cross section is
\be
\frac{d\sigma}{dE_r}(E_r,E_\nu)=\frac{G_F^2}{4\pi}Q_W^2m_N\left(1-\frac{m_N E_r}{2E_\nu^2} \right),
\ee
where we have neglected the form factor. Here $E_r$ is the recoil energy, $E_\nu$ is the neutrino energy,
$m_N$ is the mass of the target nucleus, $Q_W = A-2(1-2\sin^2\theta_W)Z$ is
the weak charge of the nucleus with $Z$ protons and $A-Z$ neutrons and $G_F$ is the Fermi-constant. The event rate is
\be
\frac{d^2R}{dE_rd\Omega_r} = \frac{\mathcal{N}}{2\pi}\Phi(t)
\int\frac{d\sigma}{dE_r}\frac{dN}{dE_\nu}\delta\left(\cos\theta_\odot-f(E_r,E_\nu)\right) dE_\nu,
\label{unint_diffrate}
\ee
\be
f(E_r,E_\nu)=\sqrt{\frac{E_r}{2m_N}}\frac{E_\nu+m_N}{E_\nu},
\ee
where $dN/dE_\nu$ is the neutrino energy distribution,
and $\mathcal{N}$ is the number of atoms per unit mass. Furthermore,
$\theta_\odot$ is the angle between the inverse solar position and the recoil momentum.
The flux $\Phi(t)$ varies with time due to changes in the Earth--Sun distance as
\be
\Phi(t)=\frac{\Phi_0}{\sqrt{1-e^2}}\frac{1}{(1-e\cos E(t))^2},
\ee
where $\Phi_0$ is the average flux on Earth, $e$ is the eccentricity of Earth's orbit,
and $E(t)$ is the eccentric anomaly at time $t$. Using the delta function to carry out
the $E_\nu$-integral yields for the differential rate
\be
\frac{d^2R}{dE_rd\Omega_r} = \frac{\Phi(t)}{2\pi}
\frac{\epsilon^2}{E_\nu^{\rm min}}\frac{d\sigma}{dE_r}(E_r,\epsilon)
\frac{dN}{dE_\nu}(\epsilon)\Theta(\cos\theta_\odot),
\label{diffrate}
\ee
where $\epsilon = (\cos\theta_\odot/E_\nu^{\rm min}-1/m_N)^{-1}$ and \mbox{$E_\nu^{\rm min} = \sqrt{m_NE_r/2}$} is
the minimum neutrino energy required to create a recoil with energy $E_r$.

The differential event rate $dR/dE_r$ is obtained by integrating the expression (\ref{diffrate})
over the solid angle using $d\Omega_r = 2\pi d\cos\theta_\odot$. For the delta-fluxes the expression can be obtained analytically,
resulting in
\be
\frac{dR}{dE_r}=\Phi(t)
\frac{d\sigma}{dE_r}(E_r,E_0)\Theta\left(1-\frac{E_0+m_N}{E_0}\sqrt{\frac{E_r}{2m_N}}\right).
\ee
The event rate as a function of the recoil energy is shown in the right panel of figure \ref{nuflux}.

\begin{figure*}
	\begin{center}
		\includegraphics[width=0.95\linewidth]{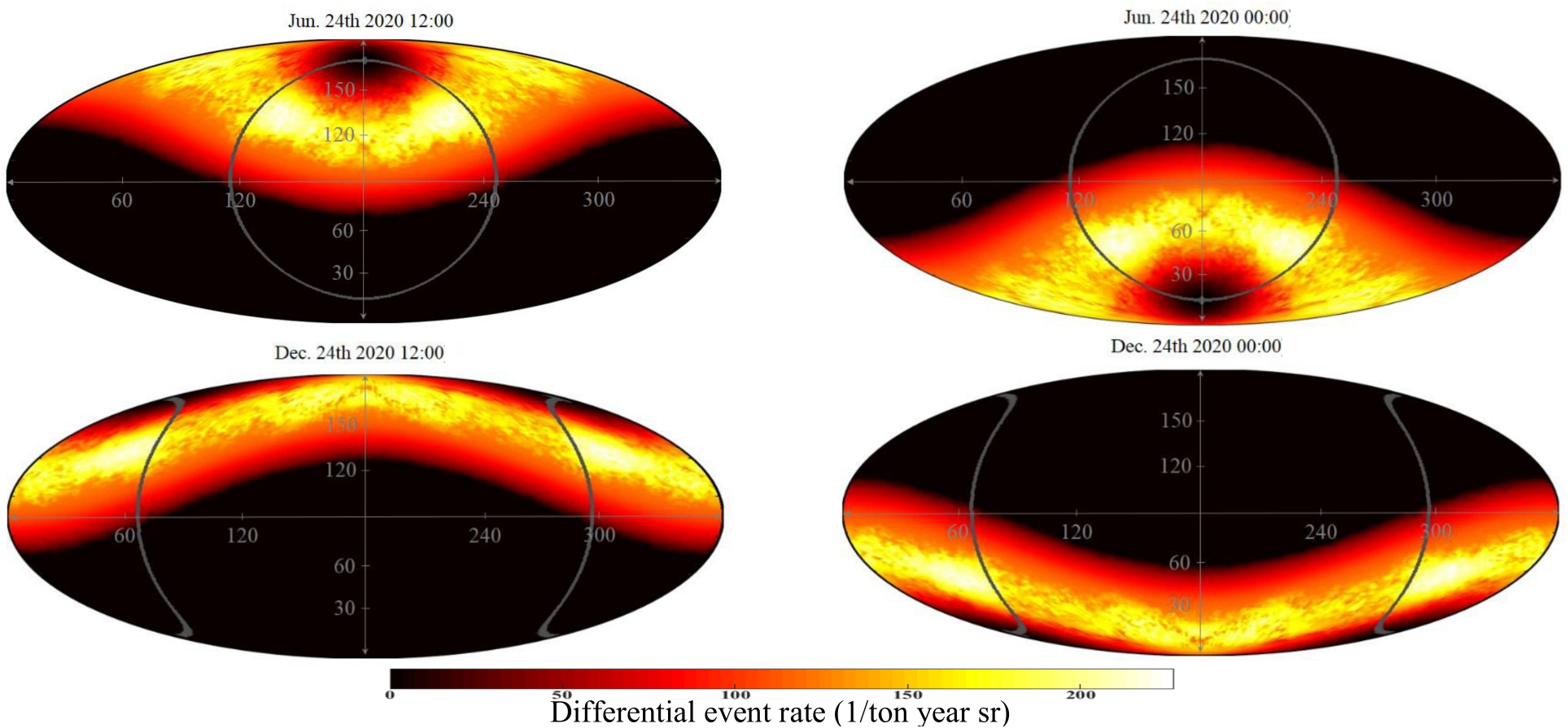}
		\caption{The nuclear recoil event rate for solar neutrinos as a function of recoil direction for germanium in June and December. The dot in each figure shows the inverse direction of the sun in the lab frame at the moment, and the loop shows how the direction varies during one day.}
		\label{neutrino_directional_rates}
	\end{center}
\end{figure*}

To obtain the event rate as a function of the recoil direction, we instead integrate the expression (\ref{diffrate}) over the recoil energy.
For the delta-fluxes this integral
can again be performed analytically, resulting in
\be
\frac{dR}{d\cos\theta_\odot}=\Phi(t)\frac{4E_0^2m_N\cos\theta_\odot}{(E_0+m_N)^2}\frac{d\sigma}{dE_r}(E_r^\theta,E_0)\Theta(E_r^\theta-E_r^{\rm min}),
\ee
where $E_r^\theta = 2m_NE_0^2\cos^2\theta_\odot/(E_0+m_N)^2$ and $E_r^{\rm min}$ is the lower limit for the energy interval of interest. The directional event rate for the solar neutrino flux is shown in figure \ref{neutrino_directional_rates}, for the same instants of time as those shown for DM in figure \ref{directional_rates}.

\begin{figure*}
\begin{center}
\includegraphics[width = .95\textwidth]{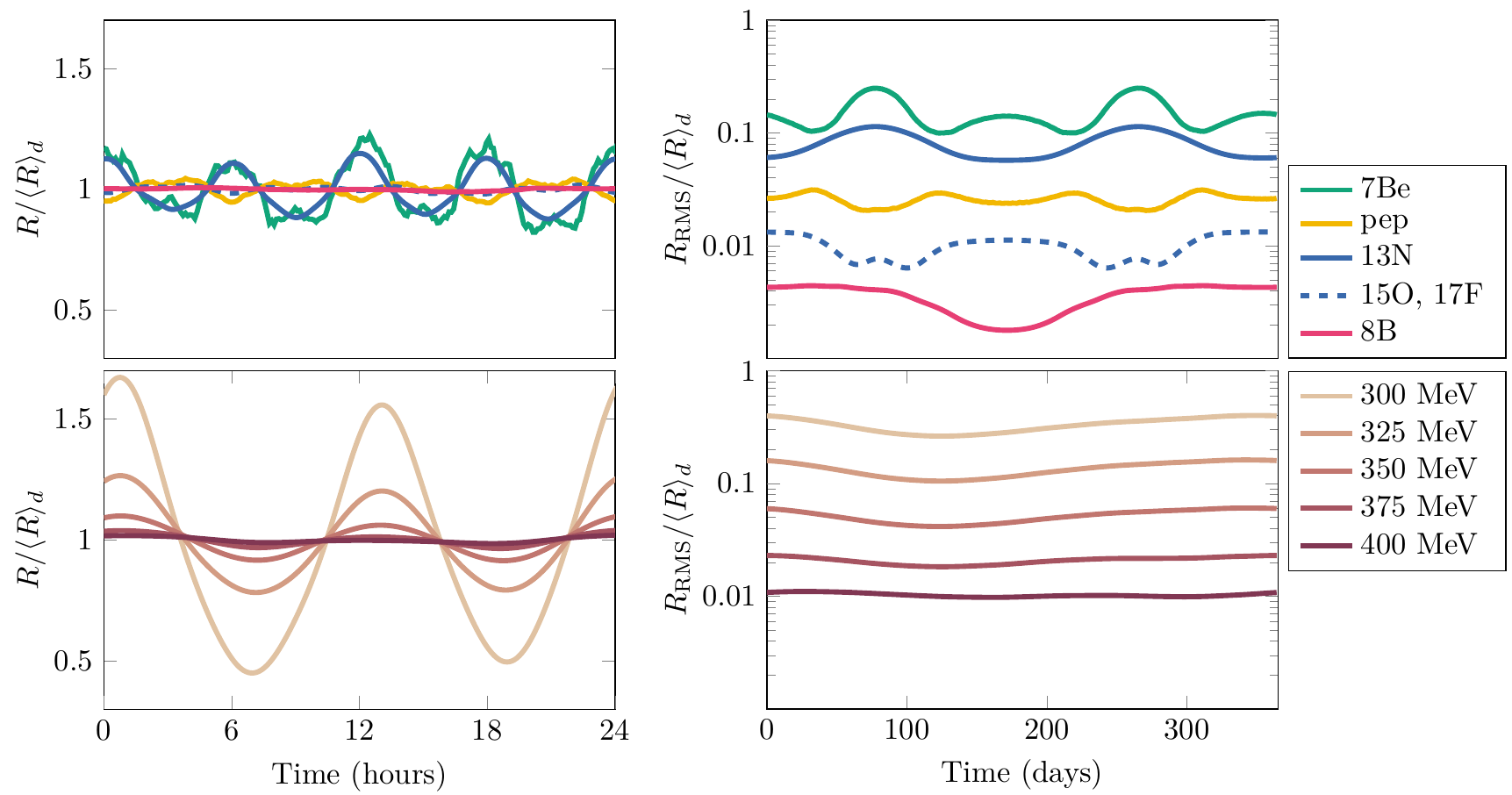}
\caption{Diurnal modulation of the event rate, i.e., the instantaneous rate divided by the average rate over one day, on germanium for solar neutrinos (upper panel) and dark matter (lower panel). The left panel shows the variation of the event rate during one day, and the right panel shows how the amplitude of the diurnal modulation varies during the year.}
\label{dailymod}
\end{center}
\end{figure*}

To obtain the instantaneous event rate in a germanium crystal we numerically integrate the directional rate $dR/d\cos\theta_\odot$ over the solid angle, by substituting the value of $E_r^{\rm min}$ from our dataset for Ge, for each direction sampled in the unit sphere. The direction of Sun appearing in the flux is a function of time, resulting in the diurnal modulation signal shown in figure \ref{dailymod}.

\section{Likelihood analysis}
\label{sec:likelihood}

To find the experimental reach of an experiment in terms of the DM-nucleon scattering cross section and the DM mass, we perform a likelihood-analysis for the identification of the DM signal from the solar neutrino background. Our treatment follows the procedure described in \cite{OHare:2015utx}\footnote{Another useful reference for the statistical analysis is \cite{Cowan:2010js}}. The general idea is that the experiment will produce data that is binned to form a histogram $(n_1,\ldots,n_N)$, where $N$ is the number of bins and $n_i$ is the number of events in bin $i$. In the following we shall consider an experiment with a given target mass that gathers data for one year, which is divided into $N_\text{bin}=N_eN_t$ bins, with $N_e=20$ being the amount of energy bins, and $N_t$ being the number of time bins, for which we consider the cases $N_t=1,365,8760$, corresponding to no time binning, daily binning that sees seasonal variation but is blind to diurnal variation in the signal, and hourly binning, which can see the diurnal variation. To perform the likelihood analysis we generate the expected event counts $n^{\nu_j}_i$ for each neutrino species ($j=$ \{pep, 7Be, 15O, 13N, 8B\}), and expected event counts $n^{\rm DM}_i$ for the DM signal, for each bin $i=1,\ldots,N_\text{bin}$.

The nuisance parameters in our analysis are the normalizations of the neutrino event rates. The neutrino fluxes and their standard deviations used in this study are given in table \ref{tab:fluxes}. The likelihood function is defined as
\begin{align}
\mathcal{L}(\mu,\{ N_j \}) =& e^{-\sum\limits_j \frac{(1-N_j)^2(\Phi^{\nu_j})^2}{2(\sigma^{\nu_j})^2}}e^{-\sum\limits_{i=1}^N\left( \mu n^{\rm DM}_i+\sum\limits_j N_j n^{\nu_j}_i\right)}\nonumber \\
&\prod\limits_{i=1}^N\frac{1}{n^{\rm obs}_i!}\left(\mu n^{\rm DM}_i+\sum\limits_j N_j n^{\nu_j}_i\right)^{n^{\rm obs}_i}.\nonumber
\end{align}
Here $n^{\rm obs}_i$ is the observed number of events in bin $i$, which in our analysis comes from a random-generated pseudo-experiment as will be explained below, $\mu$ is the normalization of the DM event rate  and $N_j$ are the relative normalizations of the neutrino fluxes, so that $N_j=1$ corresponds to the central value of the flux $\Phi^{\nu_j}$ given above. Following \cite{OHare:2015utx}, we define the likelihood ratio
\be
\lambda(0) = \frac{{\rm max}_{\{N_j\} }\left(\mathcal{L}(0,\{ N_j \})\right)}{{\rm max}_{\{ \mu,N_j\} }\left(\mathcal{L}(\mu,\{ N_j \})\right)},
\label{likelihoodratio}
\ee
where the numerator corresponds to the maximum value of the likelihood function for $\mu=0$ and the denominator for the maximum value when both $\mu$ and the normalizations $\{ N_j \}$ are allowed to vary. The test statistic $q_0$ is defined as
\be
q_0 = \begin{cases} -2\log \lambda(0)&, \hat\mu >0 \\ 0&, \hat\mu <0 \end{cases},
\ee
where $\hat\mu$ is the value of $\mu$ that maximises the denominator in (\ref{likelihoodratio}).

\begin{figure}
	\begin{center}
		\includegraphics[width=.45\textwidth]{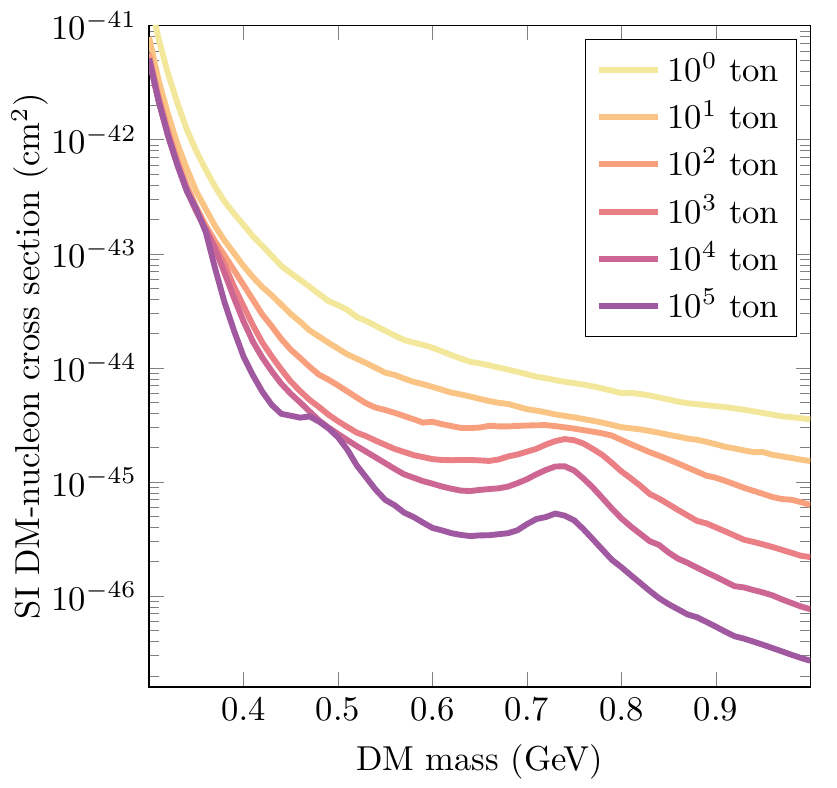}
		\caption{The reach of a germanium detector using 20 energy bins and no time information of the event rate, i.e., the neutrino floor for varying exposure of the experiment.}
		\label{fig:neutrinofloor}
	\end{center}
\end{figure}

Our procedure for finding the sensitivity of the detector for a given DM mass is as follows: First, we select a DM--nucleon cross section $\sigma^0$ and generate a random sample of the corresponding binned DM event count $n^{\rm obs,DM}_i$ in bin $i$. We do this by drawing a random number from a Poisson distribution $P(\lambda)$ with $\lambda = n^{\rm DM}_i$ for each bin $i$. This corresponds to the number of observed signal events in each bin in our simulated experiment. Next we repeat this procedure for each neutrino species, resulting in the observed background event counts $n^{{\rm obs},\nu_j}_i$ due to each neutrino species. Then the total number of observed events in bin $i$ is given by
\be
n^{\rm obs}_i = n^{\rm obs,DM}_i + \sum\limits_j n^{{\rm obs},\nu_j}_i.
\ee
We then find the values of $N_j$ that maximize $\mathcal{L}(0,\{ N_j \})$, and the values of $N_j$ and $\mu$ that maximize $\mathcal{L}(\mu,\{ N_j \})$, resulting in a value for the test statistic $q_0$. We repeat this procedure for a number of times (2000 in this case), and record the percentage of times when $q_0>9$, corresponding to the exclusion of the background-only hypothesis at three standard deviations confidence level. If that percentage is larger than 90\%, we conclude that the corresponding cross section $\sigma^0$ is within the reach of the experiment.
We then find the smallest $\sigma^0$ for which more than 90\% of the simulated experiments produce $q_0>9$, and this is the sensitivity limit (at three sigma) of the experiment for the given dark matter mass. Figure \ref{fig:neutrinofloor} shows the neutrino floor, i.e., the reach of a germanium experiment not using any time-information of the events, as a function of the exposure of the experiment.

\section{Results}
\label{sec:results}

\subsection{Discovery limits with diurnal modulation, a parametric model}

To gain insight into how different modulation signals could be discovered in the data, we consider first a simple
parametrization of the event rate given by
\be
R\sim (1+A_a\sin(\Omega t))(1+A_d\sin(\omega t)),
\label{eq:toymodel}
\ee
where $\Omega\sim 2\pi/365\,{\rm d}$ is the frequency of yearly modulation and $\omega\sim 2\pi/{\rm d}$ is the diurnal modulation frequency, and $A_a$ and $A_d$ are the amplitudes of the respective modulations. This parametrization captures the main features of the time variation of the event rate. We study the experimental reach for the time dependent signal as a function of the amplitudes $A_a$ and $A_d$. We restrict $A_a,A_d$ to the interval $[0,1]$ as required by positivity of the event rate at all times. In the dark matter direct detection setting, the amplitude $A_a$ corresponds to the well known effect of annual modulation, due to the change in the average speed of the DM wind in the laboratory frame between summer and winter, whereas the diurnal modulation amplitude $A_d$ corresponds to the effect due to crystal anisotropy.

\begin{figure}
  \begin{center}
  \includegraphics[width=.45\textwidth]{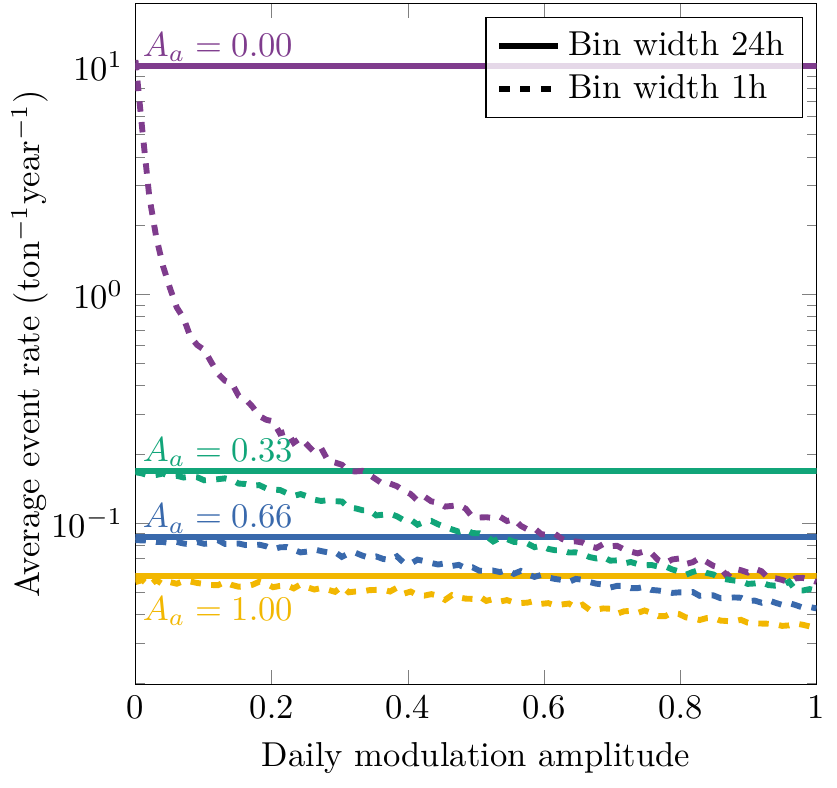}
  \includegraphics[width=.45\textwidth]{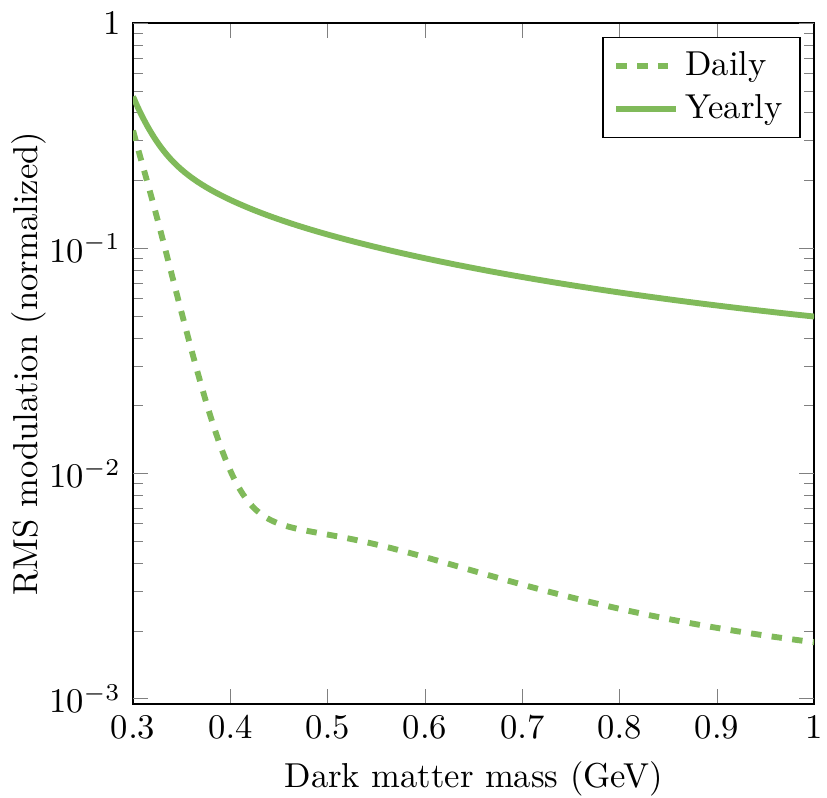}
  \caption{Top: An illustration of the gain in discovery reach due to the diurnal modulation effect on top of the annual modulation amplitude $A_a$ at exposure of $10^6$ ton years. The solid lines show the discovery reach for the parametric model using 24h bins, and the dashed lines for 1h bins, as a function of the diurnal modulation amplitude $A_d$.
  Bottom: The RMS-amplitude of the diurnal and annual modulation signals for dark matter in germanium, as a function of the DM particle mass.}
  \label{fig:example}
\end{center}
\end{figure}

The aim of this parametrization is to illuminate how the sensitivity of the detector to either modulation effect can increase the reach of the experiment, compared to a treatment where this information is not available. For this purpose we perform the likelihood analysis described in section \ref{sec:likelihood} for a signal following the parametrization (\ref{eq:toymodel}), using two different bin widths for the time bins, (a) 1h and (b) 24h. In this simplified model of modulation we assume no energy dependence of the signal or background rates. The background is modeled as a constant rate, normalized to the average solar neutrino event rate in germanium with an exposure of $10^6$ ton years. We use this unrealistically large exposure to demonstrate the behavior of the reach as a function of the model parameters. For a smaller exposure the effect is similar but smaller in magnitude. The simulated experiment (b) with the 24h bin width corresponds to an experiment that is sensitive to the annual modulation but does not observe the diurnal modulation effect, whereas the simulated experiment (a) with the 1h bin width corresponds to a detector capable of observing both effects. The results are shown in Fig~\ref{fig:example}.

We notice that, as expected, the reach of the experiment (a) (dashed lines) improves compared to the experiment (b) (solid lines) as the diurnal modulation amplitude grows. However, if the annual modulation amplitude $A_a$ is large, the additional gain due to a realistic diurnal modulation signal $A_d\lesssim 0.5$ remains modest. Conversely, the additional gain due to the annual modulation effect compared to the diurnal modulation only, can be seen as the difference between the green, blue or yellow dashed lines and the purple dashed line, for the annual modulation amplitudes $A_a$ indicated in the figure. The purple dashed line shows the effect of ignoring the annual modulation, or equivalently, a signal containing just the diurnal modulation feature. Again, if the diurnal modulation amplitude is large, the additional gain due to the annual modulation remains modest. We conclude that for a signal containing significant amplitudes for both modulation features, the information contained in each one is largely redundant with the other. In a practical experimental setup this redundancy can be used to cross check the DM origin of the signal and to control for systematic effects.

 The right panel of the figure shows how the diurnal and annual modulation amplitudes scale as a function of DM mass for a germanium target due to the defect creation threshold effect. While this behavior is specific to the target material and mechanism behind the diurnal modulation, the relative importance of the two amplitudes observed in the left panel for the parametric model holds for any diurnal modulation signal.

\subsection{Discovery limits for a germanium detector}

\begin{figure}
	\begin{center}
		\includegraphics[width=.45\textwidth]{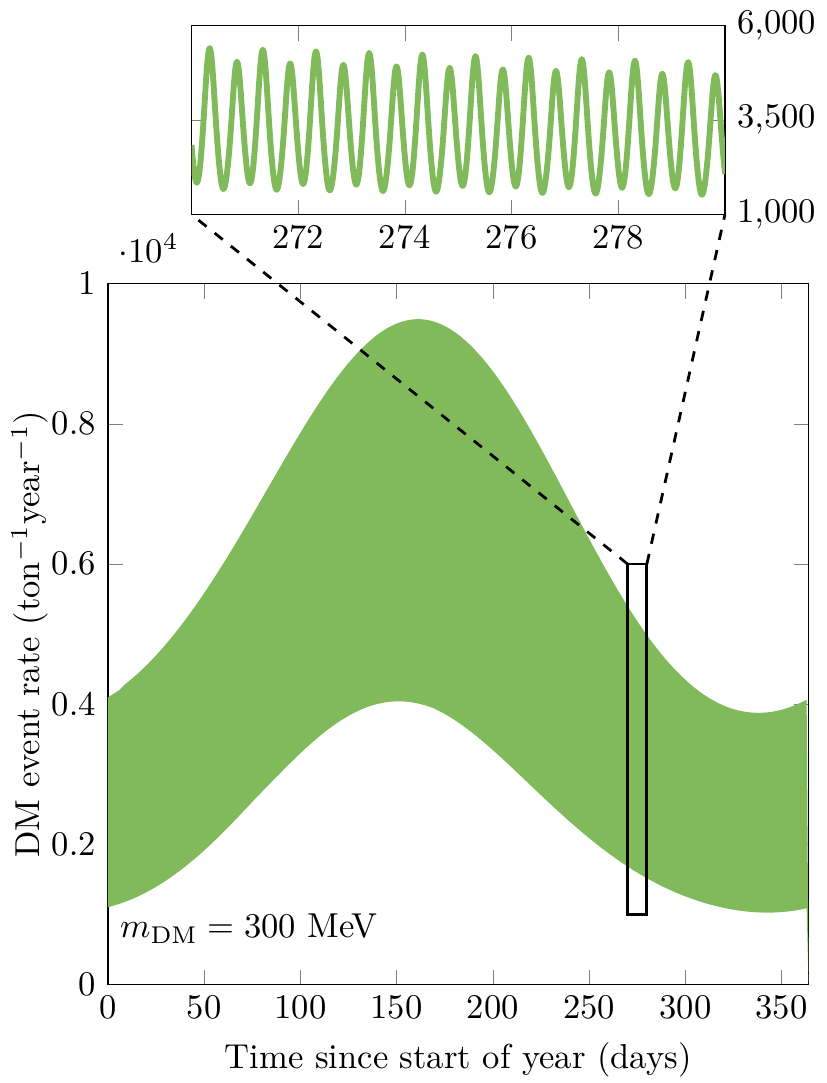}
		\caption{The event rate in germanium for a 300 MeV DM particle with a $10^{-39}\ \text{cm}^2$ cross section.}
		\label{fig:DM_modulation}
	\end{center}
\end{figure}

We now perform the likelihood analysis described in section \ref{sec:likelihood} for the DM and neutrino event rates computed with the actual threshold energy surface of germanium, as discussed in \cite{Heikinheimo:2019lwg}. To see how the actual signal event rate relates to the parametrization in (\ref{eq:toymodel}), we show the signal rate for $m_{\rm DM}=300$ MeV in figure \ref{fig:DM_modulation}. As noted in \cite{Heikinheimo:2019lwg}, this value for the DM mass results in nearly maximal diurnal modulation amplitude, corresponding to $A_d\approx 0.6$ in eq (\ref{eq:toymodel}). However, as seen in Fig~\ref{fig:DM_modulation}, this low DM mass also results in an enhanced annual modulation, corresponding to $A_a\approx0.4$ in eq (\ref{eq:toymodel}). This is because the low DM mass makes a large part of the low-speed tail of the DM velocity distribution unobservable, since the low-energy recoils remain below the detection threshold for most (or all) of the recoil directions. As the speed of the DM wind grows towards the summer, this results in a large enhancement of the event rate, because a large part of the tail of the distribution now shifts above the observation threshold.

\begin{figure*}
	\begin{center}
		\includegraphics[width=.95\textwidth]{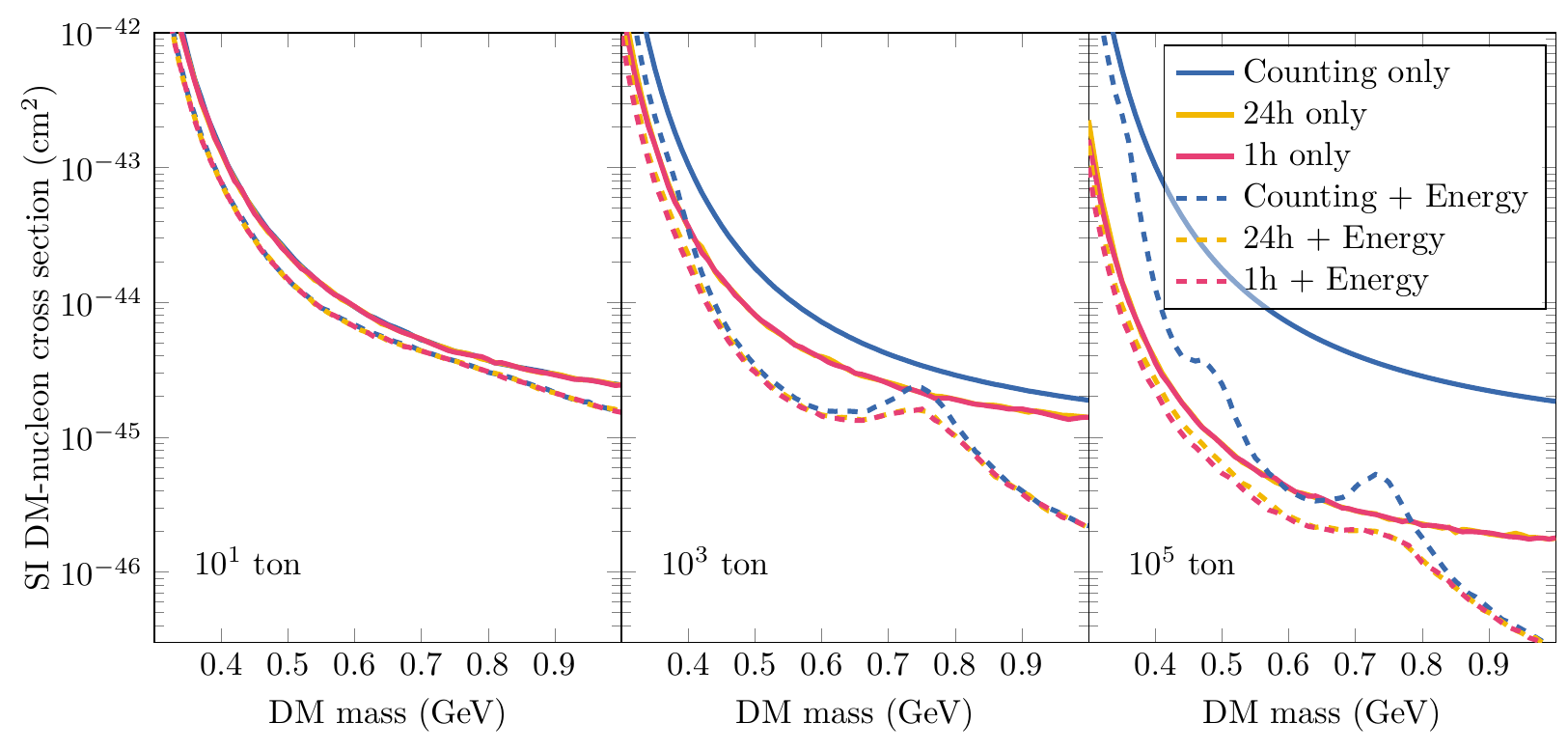}
		\caption{The discovery reach for a germanium detector with the exposure ranging from ten ton years to (an unrealistic) $10^5$ ton years. The solid lines correspond to experiments using no energy information, whereas dashed lines include energy information. The colors correspond to the resolution of time information used from no time information (blue), to 24h binning (yellow), and 1h binning (red).}
		\label{fig:discoveryreach}
	\end{center}
\end{figure*}

Since this diurnal modulation signal is ultimately a threshold effect, the dynamics described above always hold in the region of DM mass where a significant diurnal modulation is expected. This is demonstrated in the bottom panel of figure \ref{fig:example}, where the RMS-amplitude of the diurnal and annual modulation effects on the DM event rate on germanium are shown as a function of the DM mass.
 As expected based on the analysis of the parametric model above, also for the realistic event rate in germanium the gain in reach from using either modulation becomes significant for large exposure, but the additional gain from the complete time-information containing both features remains modest. The discovery reach as a function of the DM mass is shown in figure \ref{fig:discoveryreach}, where we can observe these dynamics: the blue lines that do not use any time information show a large loss of sensitivity around the values of DM mass that correspond to recoil energies of the solar neutrinos, at $\sim 500$ MeV and $\sim 750$ MeV. Making use of the annual modulation (yellow lines) almost completely removes the bump, and improves the sensitivity at best by a factor of $\sim 5$. The additional gain from using also the diurnal modulation (red line) is limited to the low-mass region $m_{\rm DM} \lesssim 500\ {\rm MeV}$, and is at best a factor of $\sim 20\%$. For comparison, the solid lines show the effect of neglecting the information contained in the recoil spectrum, i.e., using just one energy bin in the analysis, whereas the dashed lines assume perfect energy resolution, using 20 energy-bins. A real experiment with finite energy resolution will therefore obtain a reach somewhere in between the dashed and solid lines.

\subsection{Discovery limits above the neutrino floor: annually modulating background}

Finally, we explore the utility of the diurnal modulation signal in a scenario where the background exhibits an annual modulation, mimicking that of the dark matter signal. In this case the timing information from the annual modulation is obviously ineffective for separating the signal from the background, and the intra-day variation in the event rate is the only useful timing information.

Such background model might seem at first {\it{ad hoc}}, but we note that the DAMA-experiment~\cite{Bernabei:2018jrt} reports an annual modulation in their observed event rate, with the phase matching that expected from a DM signal.
However, it is very difficult to find models of DM that fit the observed modulation amplitude and energy spectrum of DAMA while being compatible with limits from other direct detection experiments, see e.g. \cite{Baum:2018ekm,Kang:2018pfq,Buckley:2019skk}.
Therefore, a reasonable doubt exists that DAMA is observing some kind of non-DM related
background with an annually modulating event rate that happens to coincide with the expected
annual modulation of a DM signal. We note that a modulation correlating with the change in
Earth--Sun distance has been reported in certain radioactive decays, see e.g.~\cite{Nistor:2013gsa}.
In any case, whatever the mechanism responsible for the annual
modulation observed in DAMA, it is clear that observing the diurnal modulation signal would provide an independent confirmation of the DM-origin of the events.

As our study here is focused on the germanium detector, our results are not directly relevant for the NaI target of DAMA. We plan to study the utility of the diurnal modulation feature in the NaI target in future work, and report here the potential gain in sensitivity of a germanium detector from using the intra-day time information of the event rate, in the case that the background perfectly imitates the annual modulation of DM.

\begin{figure}
	\begin{center}
		\includegraphics[width=.45\textwidth]{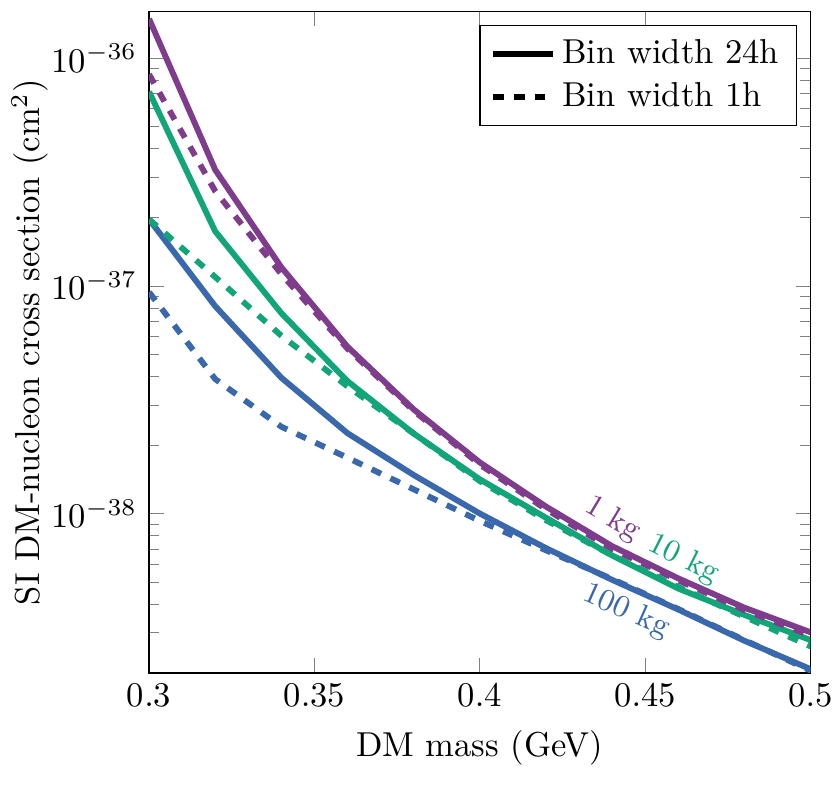}
		\caption{The discovery reach for a germanium detector with the exposure ranging from one kilogram year to 100 kilogram years. The dashed lines correspond to the experiment using the full time information (1h), while the solid lines show the reach when averaging over one day. The background rate is time dependent, matching the annual modulation of the DM signal, normalized to $10^5$ events per kg year.}
		\label{fig:impostor}
	\end{center}
\end{figure}

Figure \ref{fig:impostor} shows the reach for the germanium experiment with a mass from one kilogram to 100 kilograms with 24-hour bins (solid lines), and one-hour bins (dashed lines).
The gain in sensitivity reach from using the diurnal modulation can be seen as the difference between the dashed and solid lines. For a 300 MeV DM particle we observe a gain by a factor of $\gtrsim 2$.

\section{Conclusions and outlook}
\label{sec:checkout}

We conclude that within the parameter space where a significant diurnal modulation of the DM signal is expected, the same threshold effect that gives rise to the diurnal modulation also results in an enhanced annual modulation of the event rate. Therefore the gain in sensitivity of the experiment from time information of the recoil events
is almost saturated by observing either of the modulation signatures, and only a modest additional gain can be achieved by observing both. We anticipate that in a practical experimental situation this redundancy may prove very useful in cross checking the results and for controlling systematics. 

Furthermore, as discussed in \cite{Heikinheimo:2019lwg}, the structure of the diurnal modulation signal depends on the type of the DM-nucleon scattering operator, and also on the DM velocity distribution (see e.g.~\cite{Ge:2020yuf}). Here we have assumed the simplest spin-independent scattering operator and the standard halo model for DM velocity.
Observing the diurnal modulation feature would provide valuable information about the type of the scattering operator and the DM velocity distribution, and therefore about the nature of the DM particle. To reliably obtain this information, i.e., to be able to exclude a hypothesis of a given operator in favor of another, will require a larger exposure than what is needed to simply exclude the background-only hypothesis, which has been the focus of this study. We shall explore these possibilities further in our upcoming work. Finally, we note that the annual modulation reported by the DAMA experiment leaves open the possibility that an unknown background process exhibits an annual modulation imitating the DM signal. In this case the diurnal modulation effect could produce the additional information required to isolate the signal from the background. A dedicated NaI-target analysis is needed in order to find the exact consequences of this phenomenon for the DAMA-signal.

\end{document}